\documentstyle[12pt,aasms4]{article}

\newcommand{\de}{\delta}
\newcommand{\th}{\theta}
\newcommand{\al}{\alpha}
\newcommand{\si}{\sigma}
\newcommand{\bx}{{\bf x}}

\newcommand{\lan}{\langle}
\newcommand{\ran}{\rangle}

\newcommand{\be}{\begin{equation}}
\newcommand{\ee}{\end{equation}}
\newcommand{\bea}{\begin{eqnarray}}
\newcommand{\eea}{\end{eqnarray}}

\newcommand{\bef}{\begin{figure}}
\newcommand{\eef}{\end{figure}}

\def\spose#1{\hbox to 0pt{#1\hss}} 
\def\ltapprox{\mathrel{\spose{\lower 3pt\hbox{$\mathchar"218$}} 
 \raise 2.0pt\hbox{$\mathchar"13C$}}} 
\def\gtapprox{\mathrel{\spose{\lower 3pt\hbox{$\mathchar"218$}} 
 \raise 2.0pt\hbox{$\mathchar"13E$}}} 
\def\inapprox{\mathrel{\spose{\lower 3pt\hbox{$\mathchar"218$}} 
 \raise 2.0pt\hbox{$\mathchar"232$}}} 
 
 \slugcomment{Subimitted to Astrophys. J. Letters}

\lefthead{Gabrielli, Sylos Labini and Durrer}
\righthead{Biasing in Gaussian random fields and   galaxy correlations }

\begin{document} 
 
\title{Biasing in Gaussian random fields and   galaxy correlations}
\author{Andrea Gabrielli\altaffilmark{1,2}, 
Francesco Sylos Labini\altaffilmark{2,3}  
and  Ruth Durrer \altaffilmark{3}  
}

\altaffiltext{1}{Laboratoire de Physique de la 
Mati\`ere Condens\'ee, \'Ecole Polytechnique, 91128 - Palaiseau Cedex,
 France}

\altaffiltext{2}{INFM Sezione Roma1,        
		      Dip. di Fisica, Universit\'a "La Sapienza", 
		      P.le A. Moro, 2,  
        	      I-00185 Roma, Italy. } 
        	      
 \altaffiltext{3}{D\'epartement de Physique Th\'eorique, 
Universit\'e de Gen\`eve,
24 quai Ernest Ansermet, CH-1211 Gen\`eve 4, Switzerland}

\begin{abstract}

In this letter we show that in a Gaussian random field the correlation
length, the typical size of correlated structures, does not change 
with biasing.
We interpret the amplification of the correlation functions of 
subsets identified by different thresholds being due to the increasing
sparseness of peaks over threshold.   
This clarifies an long-standing misconception in the
literature. We also argue that this effect does not explain the observed 
increase of the amplitude of the correlation function $\xi(r)$
when galaxies of brighter luminosity or galaxy clusters of increasing
richness are considered.   
\end{abstract}


\setcounter{footnote}{0}
We first  explain, in mathematical terms, the notion of biasing
for a Gaussian random field. Here we follow the ideas of Kaiser (1984
which have been developed further in Bardeen et al., 1986).
We then calculate biasing for some examples and we clarify the
physical meaning of bias in the context of  Kaiser (1984).
Finally, we comment on the significance of
our findings for the correlations of galaxies and clusters.

We consider a homogeneous, isotropic and correlated continuous Gaussian 
random field,  $\de(\bx)$, with mean zero and
variance $\si^2=\lan \de(\bx)^2\ran$ in a volume $V$.
The application of the following discussion to a discrete set of points
is straightforward considering the effect of a smoothing length.
The marginal one-point probability density function of $\de$  is 
\[ P(\de) = {1\over \sqrt{2\pi}\si} e^{-{\de^2\over 2\si^2}} ~.\]
Using $P$, we calculate the  fraction of the 
volume $V$ with $\de(\bx)\ge \nu\si$,
$ P_1(\nu) = \int_{\nu\si}^\infty P(\de) d\de$.

The correlation function between two values of $\de(\bx)$ in two points
separated by a distance $r$ is given by 
$\xi(r)=\lan \de(\bx)\de(\bx+ r{\bf n})\ran$. By definition, $\xi(0)=\si^2$.
In this context, homogeneity means that the variance, $\si^2$, and the
correlation function, $\xi(r)$, do
not depend on $\bx$. Isotropy means that $\xi(r)$ does not depend on
the direction {\bf n}\footnote{In other words, we assume  
$\de(\bx)$ to be a so called
`stationary normal stochastic process' (Feller 1965).}. 
An important application we have in mind are cosmological  
density fluctuations, $\de(\bx) =
(\rho(\bx)-\rho_0)/\rho_0$, where $\rho_0 =\lan\rho\ran$ is the mean density;
but the following arguments are completely general.\footnote{Clearly,
cosmological  density fluctuations can never be perfectly Gaussian
since $\rho(\bx)\ge 0$ and thus $\de(\bx)\ge -1$, but, for small
fluctuations, a Gaussian can be a good approximation. Furthermore, our
results remain at least qualitatively correct also in the non-Gaussian
case.} 
Here and in what follows we assume that 
the average density $\rho_0$ is a well defined positive quantity.
This is not so if the distribution is fractal (Pietronero, 1987).

Our goal is, to determine the correlation function of local maxima from the
correlation function of the underlying density field. 
Like Kaiser (1984) we simplify the problem by computing the
correlations of {\it regions} above a certain threshold $\nu\sigma$
instead of the correlations of  {\it maxima}. However, these quantities are
closely related for values of $\nu$ significantly larger than 1.
We define the  threshold density, $\th_\nu(\bx)$ by
\be
 \th_\nu(\bx) \equiv \th(\de(\bx)-\nu\si)= \left\{ \begin{array}{ll}
	1 & \mbox{if }~~~ \de(\bx) \ge \nu\si \\
	0 & \mbox{else.}
\end{array} \right.
\ee
Note the qualitative difference between $\de$ which is a weighted
density field, and $\th_\nu$ which just defines a set, all
points having equal weight.
We note the following simple facts 
concerning the threshold density, $\th_\nu$, due only to its definition, 
independently on the correlation properties of $\de(\bx)$:
\bea
& &\lan \th_\nu\ran \equiv P_1(\nu)  \le  1~,~~~ 
 (\th_\nu(\bx))^n = \th_\nu(\bx)~,  \label{Eq2} \\ 
& &\lan\th_\nu(\bx)\th_\nu(\bx + r{\bf n})\ran  \le   P_1(\nu)~,
\nonumber \\ 
& &  {\lan\th_\nu(\bx)\th_\nu(\bx + r{\bf n})\ran\over P_1(\nu)^2} -1
    \equiv  
    \xi_\nu(r) \le \xi_\nu(0) = {1\over P_1(\nu)} -1~, \nonumber \\
& & \th_{\nu'}(\bx) <   \th_{\nu}(\bx) ~,~~~
P_1(\nu') \! <   \!  P_1(\nu) ~~ \mbox{ for } ~~ \nu' >
	\nu ~, \nonumber \\ 
& & \xi_{\nu'}(0)  >   \xi_{\nu}(0)  ~ \mbox{ for } ~~ \nu' >
	\nu ~.   
\label{xineq}
\eea
The difference between  $\th_\nu$ for different values of
$\nu$ is called biasing.
The enhancement of  $\xi_{\nu}(0)$ for higher thresholds has
clearly nothing to do with  
how 'strongly clustered'  the peaks are but is
entirely due to the fact that   the larger  $\nu$ the lower
the fraction of points above the threshold 
({\it i.e.} $P_1(\nu')<P_1(\nu)$ for $\nu'>\nu$).
If we consider the trivial case of white Gaussian noise 
($\xi(r) = 0 \; \mbox{for} \; r>0$) the peaks are just spikes.
When a threshold $\nu\sigma$ is considered the number of spikes 
decreases and hence $\xi_{\nu}(0)$ is amplified 
because they are much more sparse and not because 
they are `more strongly clustered': we show in the following that also
in the case of a correlated field ($\xi(r)\not{\hspace{-5pt} \equiv} 0$
for $r>0$) the importance of sparseness is crucial in order to 
explain the amplification of $\xi_{\nu}(r)$.

In the context of cosmological density fluctuations,
if the average density of matter is a well-defined positive constant,
 the amplitude of $\xi_m(r)$ of matter distribution 
is very important, since 
its integral over a given radius is
proportional to the over density on this scale,
\be
\si(R) =3R^{-3}\int_0^R\xi_m(r)r^2dr ~.
\label{R_l}
\ee
The scale $R_l$ where $\si(R_l)\sim 1$, separates large, non-linear
fluctuations, from small ones (Gaite et al., 1999).
It is very important to stress the following point: from the knowledge
of the functions $\xi_{\nu}(r)$ for two different subsets of the density field 
obtained from two different values $\nu$ and $\nu'$ of the threshold,
it is not possible to
predict the amplitude of the fluctuations of the original density 
field at any scale if we do not know the underlying values
$\nu$, $\nu'$ and $\sigma$.
On the other hand, as we are going to show, the only feature of the original field
which can be inferred by the behavior of  $\xi_{\nu}(r)$
is the large scale behavior of the correlation function $\xi(r)$,
in particular the correlation length (if this length is finite, in the
statistical physics terminology.)
The correlation length $r_c$ can be defined as (Gaite et al. 1999):
\be
r^2_c=\frac{1}{2}\left|\frac{\nabla^2 P(k)}{P(k)}\right|_{k=0}\,,
\label{def-rc}
\ee
where $P(k)$ is the Fourier transform of $\xi(r)$.
Note that  $r_c$ is independent of any multiplying constant in $\xi(r)$,
so it is not related to its amplitude. 
This correlation length is that used
in statistical physics and field theory (Ma 1984), and gives the length
scale beyond which $\xi(r)$ decays rapidly to zero (e.g. exponentially).
Roughly, this implies that the fluctuations of the field are
 organized in structures 
up to a scale $r_c$ (Gaite et al.  1999).  
However, in cosmology the correlation length has been defined historically 
(Peebles 1980)
through the amplitude of
 $\xi(r)$ by looking at 
 the distance $r_0$ at which it is equal to $1$.
Provided that a constant 
positive density $\rho_0$
 of the field exists, $r_0$ gives 
the scale beyond which the
 fluctuations becomes small with respect to $\rho_0$ (then it is
analogous to the previously 
defined $R_l$), and hence it provides also the minimal
size of a sample of the field giving a good estimate of the intrinsic $\rho_0$.  
The confusion between $r_c$ 
and $r_0$ (see also Gaite et al. 1999) 
is at the basis of the misinterpretation of the 
concept of bias, as we are going to show.   

The joint two-point  probability density
${\cal P}_2(\de,\de';r)$ depends
on the distance $r$ between $\bx$ and $\bx'$, where
$\de=\de(\bx)$ and $\de'=\de(\bx')$. 
For Gaussian Fields  ${\cal P}_2$ 
  is entirely  determined by the 2-point correlation function $\xi(r)$ 
 (Rise 1954, Feller 1965):
\bea
 && {\cal P}_2(\de,\de';r) = \\
&&={1\over 2\pi\sqrt{\si^4-\xi(r)^2}}
  \exp\left(
  -{\si^2(\de^2+\de'^2) -2\xi(r)\de\de' \over
2(\si^4-\xi^2(r))}\right) ~.
\nonumber
\eea
By definition
\be
\xi(r) \equiv \lan\de (\bx +r{\bf n})\de(\bx)\ran =
    \int_{-\infty}^\infty\int_{-\infty}^\infty d\de d\de' \de\de'
    {\cal P}_2(\de,\de';r) ~. \label{P2}
\ee 
The probability that both, $\de$ and $\de'$ are larger than
$\nu\si$ is
\be
 P_2(\nu,r) =  \int_{\nu\si}^\infty \int_{\nu\si}^\infty {\cal P}_2(\de,\de',r)
	d\de d\de'\equiv \lan\th_\nu(\bx)\th_\nu(\bx + r{\bf n})\ran~.
\ee
The conditional probability 
that $\delta(\bf y)\ge \nu\si$, 
given $\de(\bf x)\ge \nu\si$, where $|{\bf x- y}| = r$,
is then just $P_2(\nu,r)/P_1(\nu)$. 
The two-point
correlation function for the stochastic variable $ \th_\nu(\bx)$, 
introduced above can be expressed in terms of $P_1$ and $P_2$ by
\be
\xi_\nu(r) = {P_2(\nu,r) \over P_1^2(\nu)} -1 
\label{xinu1}
\ee
Defining $\xi_c(r) = \xi(r)/\si^2$, we obtain 
\bea
 P_1(\nu)^2 (\xi_\nu(r) + 1) =  {1\over 2\pi\sqrt{1-\xi_c^2}} 
	\int_\nu^\infty \int_\nu^\infty	d\! x d\! x'
\nonumber  \\
\times \exp\left(-{(x^2+x'^2)-2\xi_c(r)xx'\over 2(1-\xi_c^2(r))}\right)
\label{xinuex} 
\eea
It is worth  noting that the amplitude of $\xi_\nu(r)$ does not
 give information about how large 
the fluctuations are with respect to $\rho_0$, but it rather describes
the ``fluctuations of the fluctuations'', that is the fluctuations of
the new variable $\th_\nu(\bx)$ around its average $P_1(\nu)$.
Similar arguments to those introduced for the original field can now be
developed to characterize the typical scales of the new set defined by 
$\th_\nu (\bx)$. In particular, one can define a correlation length
$r_c(\nu)$ using the analog of Eq.~(\ref{def-rc}), by replacing $\xi(r)$
with $\xi_\nu(r)$. Like $r_c$,
$r_c(\nu)$ does not depend on any multiplicative
constant in $\xi_\nu(r)$, i.e. it does not depend on the amplitude of 
$\xi_\nu(r)$.  Moreover a 'homogeneity scale' $r_0(\nu)$ can be
defined looking at the scale 
at which $\xi_\nu(r)=1$ (or alternatively Eq.~\ref{R_l}). The value of 
$r_0(\nu)$ strongly depends on the amplitude of $\xi_\nu(r)$ and 
represents the minimal size of a sample of the set
giving meaningful estimates of the average density $P_1(\nu)$ and of $r_0(\nu)$
itself; $r_0(\nu)$ is the distance at which the conditional density $P_2(\nu,r)/P_1(\nu)$ 
begins to flatten towards $P_1(\nu)$.
We show below that while $r_0(\nu)$ depends strongly on $\nu$ due to a sparseness effect, 
$r_c(\nu)$ is almost constant and equal to $r_c$ of the field, i.e. the maximal size
of the fluctuations' structures does depend on the threshold. 

Eq.~(\ref{xinuex}) implies, for $\nu\gg 1$ and for sufficiently large $r$  such
that $\xi_c(r)\ll 1$  ( Politzer \& Wise, 1984):
\be
\xi_\nu(r) \simeq \exp\left({\nu^2\xi_c(r)}\right) -1 ~,
\label{xinu}
\ee
to lowest non-vanishing order in $\xi_c(r)$. If, in addition,
$\nu^2\xi_c(r)\ll 1$ we find (Politzer \& Wise, 1984)
\be 
\xi_\nu(r) \simeq {\nu^2}\xi_c(r) ~. \label{xinuapp}
\ee
This is the relation derived by Kaiser (1984). He only states
the condition $\xi_c(r) \ll 1$ and separately $\nu\gg 1$, which is
 significantly weaker than the required 
$\nu^2\xi_c(r) \simeq  \xi_\nu(r) \ll 1$, especially around the 
correlation length where $\xi$ is not yet very small.

It is important to note that in the cosmologically relevant regime,
$\xi_\nu \gtapprox 1$ the Kaiser relation 
(eq.\ref{xinuapp}) does not apply
and $\xi_\nu$ is actually exponentially enhanced. If this mechanism
would be the cause for the observed cluster correlation function one
would thus expect an exponential enhancement on scales where 
$\xi_{cc} \gtapprox 1$, {\em i.e.} $R\ltapprox 20h^{-1}$Mpc. This is in
contradiction with observations (Bahcall \& Soneira  1983)!\footnote{One
might argue that non-linearities which are important when the
fluctuations are large can ``rescue'' the Kaiser
relation~(\ref{xinuapp}) also into the regime $\xi_\nu>1$. There are
two objections against this: First of all, as we pointed out above,
$\xi_\nu>1$ does not imply large fluctuations of the original density
field. Actually most
cosmologists would agree that on $R\sim 20h^{-1}$Mpc, where the
cluster correlation function, $\xi_{cc}\sim 1$, fluctuations are
linear. Secondly, it seems very unphysical that Newtonian clustering
should act as to change the exponential relation~(\ref{xinu}) into a
linear one~(\ref{xinuapp}).}

If, within a  range of scales, $\xi(r)$ can be approximated by
a power law, $\xi =(r / r_0)^{-\gamma}$, and if the threshold $\nu$ is
such that Eq.~(\ref{xinuapp}) holds,   which implies 
$\xi_\nu\ll 1$, we have $ \xi_\nu =(r/r_0(\nu))^{-\gamma}$.
The scales $r_0(\nu)$ for different biases are related by
$ r_0(\nu') = r_0(\nu)(\nu'/\nu)^{2/\gamma} ~. $
For that reason Kaiser, who first derived relation~(\ref{xinuapp}),
interpreted it as an increase in the ``correlation
length'' $r_0(\nu)$, which in our language is the homogeneity scale of the set 
$\th_\nu(\bx)$.

In order to clarify the meaning of the two length 
scales $r_c(\nu)$ and $r_0(\nu)$
we first study an example of a Gaussian density field with
finite correlation length $r_c$, and which is well approximated by a
power law on a certain range of scales. 
The case in which $r_c \rightarrow \infty$ is straightforward.
We set
\[\xi(r) =\frac{\sigma^2\exp(-r/r_c)}
{ [1+(k_s r)^\gamma]} \] with $k_s^{-1}\ll r_c$;  
$k_s^{-1}$ represents the smoothing scale of the continuous field, 
which is characterized better in the following, and $r_c$ is approximately
the correlation length as defined as Eq.~(\ref{def-rc}).
In the region $k_s^{-1}\ll r \ll r_c$, $\;\;\xi(r)$  is well
approximated by the power law $(k_s r)^{-\gamma}$. 
The correlation lengths, $r_c(\nu)$ for any value of $\nu$, are given by the
slope of $\log\xi_\nu(r)$ at large $r$, vs. $r$, which is clearly
independent of bias  (Fig.\ref{fig1}).
This can also be obtained from Eqs.~(\ref{xinu},\ref{xinuapp}).
\placefigure{fig1} 

For relatively small values of the threshold, 
$\nu \ll \nu_c \approx (k_sr_c)^{\gamma/2}$ one finds in this case
$r_0(\nu) \ll r_c$ and $r_0(\nu) \sim k_s^{-1} \nu$. On the other hand, 
if  $\nu \gg \nu_c$ we have $r_0(\nu) \sim r_c \log(\nu)$
and in this case the statistics is dominated
by shot noise (see below). For this reason  we assume
$r_0(\nu)<r_c(\nu)$ in the following.
We note that in the range of scales $r \le r_0(\nu)$ the amplification
of $\xi_\nu(r)$ is strongly non linear in $\nu$ and it is scale dependent:
hence if the original correlation function $\xi(r)$ has a power law behavior,
$\xi_{\nu}(r)$ does not for $r \le r_0(\nu)$:
this is better shown in the case in which $r_c \rightarrow \infty$.
In this case the correlation function is 
\be
 \xi(r) = \frac{\si^2}{(1+(r k_s^{-1})^\gamma)} ~.
\ee 
Clearly on scales $k_s^{-1} <r <r_c$ this example does not differ 
from the above (but of course the correlation length is infinite here).
The amplification of $\xi_\nu$ for this example is plotted in
Fig.~\ref{fig2}.
In order to investigate whether $\xi_\nu(r)$ is of the form $\xi_\nu(r)\sim
(r/r_0(\nu))^{-\gamma_\nu}$, we plot $-d\log(\xi_\nu(r))/d\log(r) \sim
\gamma_\nu$ in Fig~\ref{fig3}. Only in
the regime where $\xi_\nu(r)\ll 1$, $\gamma_\nu$ becomes constant and roughly
independent of $\nu$. This behavior is very different from the
one  found in galaxy catalogs!

\placefigure{fig2} 

\placefigure{fig3} 

Let us now clarify how the amplification of $\xi_\nu(r)$ is related to
the increase of the peak sparseness  with the threshold $\nu$.
 For a Gaussian random field, the mean peak size,
$D_p(\nu)$ and the mean peak distance, $L_p$ are respectively (Vanmarcke
1983, Coles 1986):
$D_p(\nu) \simeq D_0(k_s,r_c)/ \nu$ and
    $L_p(\nu) \simeq D_0(k_s,r_c)\exp(\nu^2/6)\nu^{-2/3} $
so that 
\be L_p/D_p \simeq \nu^{1/3}\exp(\nu^2/6) 
\mbox{ for } \nu\gg 1~, 
\label{peaks}
\ee
$D_0(k_s,r_c)$ is given by 
\be
D_0^2={\int_{0}^{+\infty} dkP_1(k) \over \int_{0}^{+\infty} 
dk k^2  P_1(k)}
\ee
where $ P_1(k)$ is the Fourier transform of $\xi(r)$ along a line in space
(in $d=1$ it coincides with $P(k)$). Eq.~(\ref{peaks}) shows the
strong enhancement of the sparseness of peaks (object) with 
increasing $\nu$. It is this  increase of sparseness which is at the 
origin of the amplification by biasing.
In the light of Eqs.~(\ref{xinu},\ref{xinuapp},\ref{peaks}), we see
that  increasing $\nu$ corresponds to a very particular sampling of 
fluctuations: the typical size of the surviving peaks $D_p$ is slowly 
varying with $\nu$ while the average distance between peaks $L_p$
is more than exponentially amplified, and finally the scale
$r_c(\nu)$, over which the fluctuations are structured, is practically
unchanged. 
 
We have argued that  bias does not influence the
correlation length ($r_c(\nu)\simeq r_c$). It amplifies the
correlation function by the fact that the mean density,   $P_1(\nu)$,  
is reduced more strongly than the conditional density, $P_2(\nu,r)/P_1(\nu)$. 
According to Eq.~(\ref{xinu}), this amplification is strongly
non-linear in $\xi(r)$ (exponential) at scales where $\nu^2
\xi_c(r)\ge 1$  and thus $\xi_\nu(r)>1$.

Consequently, as we want to stress once more, the biasing mechanism
introduced by Kaiser and discussed in this work
cannot lead to a relation of the form 
$\xi_{\nu'}(r) =\al_{\nu'\nu}\xi_{\nu}(r)$
over a range of scales $r_1<r<r_2$ such that $1<\xi_{\nu}(r_1)$ and
$\xi_{\nu}(r_2)<1$. But exactly this behavior is found in galaxy and cluster
catalogs. For example in (Bahcall \& Soneira  1983 ) or
 (Benoist et al. 1986), a constant 
biasing factor $\al_{\nu'\nu}$ over a range from about $1h^{-1}$Mpc  
to $20h^{-1}$Mpc is
observed for correlation amplitudes varying from about $20$ to 
$0.1$. We therefore conclude that the
explanation by Kaiser (1984) cannot be at the origin of the
difference of the correlation functions
observed in the distribution of galaxies with different intrinsic
magnitude or in the distribution of clusters with different richness. 
 
This result appears at first disappointing since it invalidates an
explanation without proposing a new one. On the other hand, the search
for an explanation of an observed phenomenon is only motivated if we
are fully aware of the fact that we don't already have one. 

Last but not least, we want to point out that fractal density fluctuations
together with the fact that more luminous objects are seen out to
larger distances do actually induce a 
increase in the amplitude of the correlation 
function $\xi(r)$ similar to the one observed in real galaxy 
catalogs (Pietronero 1987, Sylos Labini,
Montuori \& Pietronero 1998). In this explanation, the linear   
amplification found for the correlation function, has 
nothing to do with a correlation length but is a pure finite size effect,
and the distribution of galaxies does not have any intrinsic characteristic
scale.

 It is a pleasure to thank 
  A. Baldassarri,
  P. Ferreira, M. Montuori 
  and 
  L. Pietronero,
 for useful
 discussions.
 This work is partially supported by the 
 EEC TMR Network  "Fractal structures and  self-organization"  
\mbox{ERBFMRXCT980183} and by the Swiss NSF.

\clearpage

\figcaption[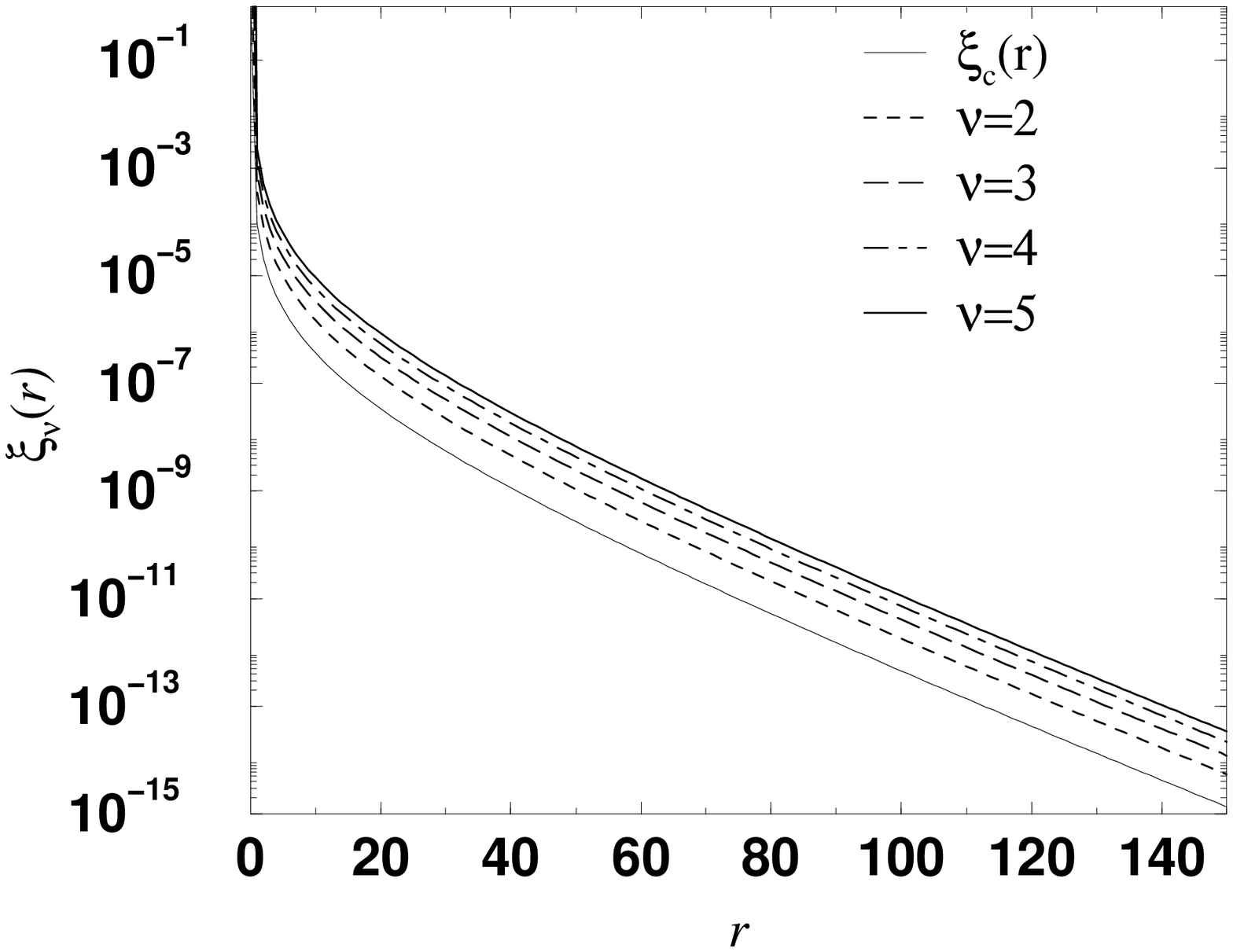]
{\label{fig1}
Behavior of
$\xi(r) = \sigma^2 / [1+(k_s r)^\gamma \exp(-r/r_c)$
(where $\gamma=-2$, $k_s^{-1}=0.01$ and $r_c=10$)
 and $\xi_{\nu}(r)$ are shown for different
values of the threshold $\nu$ in a semi-log plot.
The slope of $\xi_{\nu}(r)$ for $r\gtapprox 50$ is $-1/r_c$,  independent of $\nu$
{\em i.e.} the correlation length of the system does not
change for the sets above the threshold.
}

\figcaption[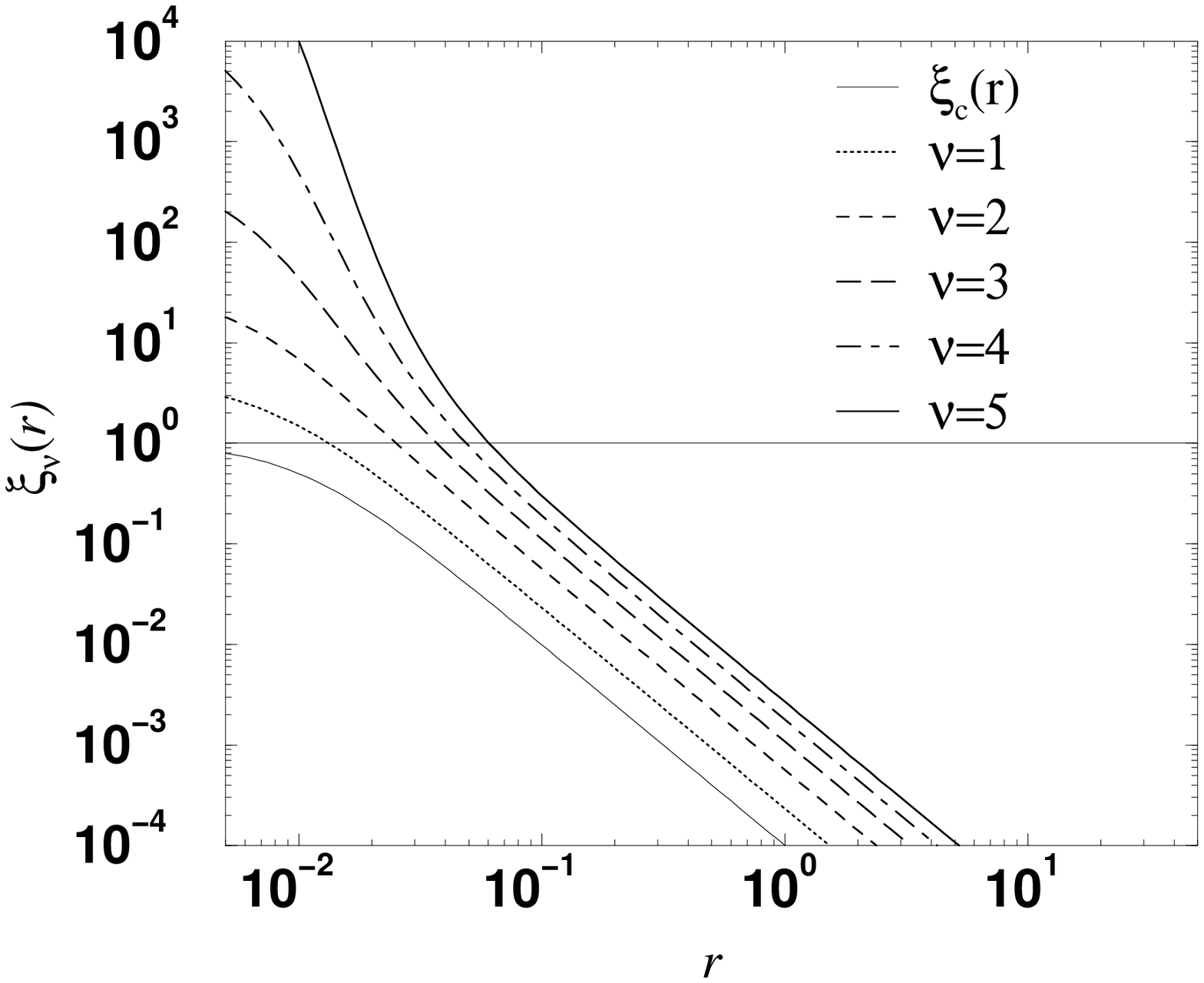]
{\label{fig2} 
 Behavior of
$\xi(r) \sim \sigma^2/(1+ (k_s r)^{\gamma})$
(with $\gamma=-2$, $k_s^{-1}=0.01$)
 and $\xi_{\nu}(r)$ are shown for different
values of the threshold $\nu$ in a log-log plot.
}

\figcaption[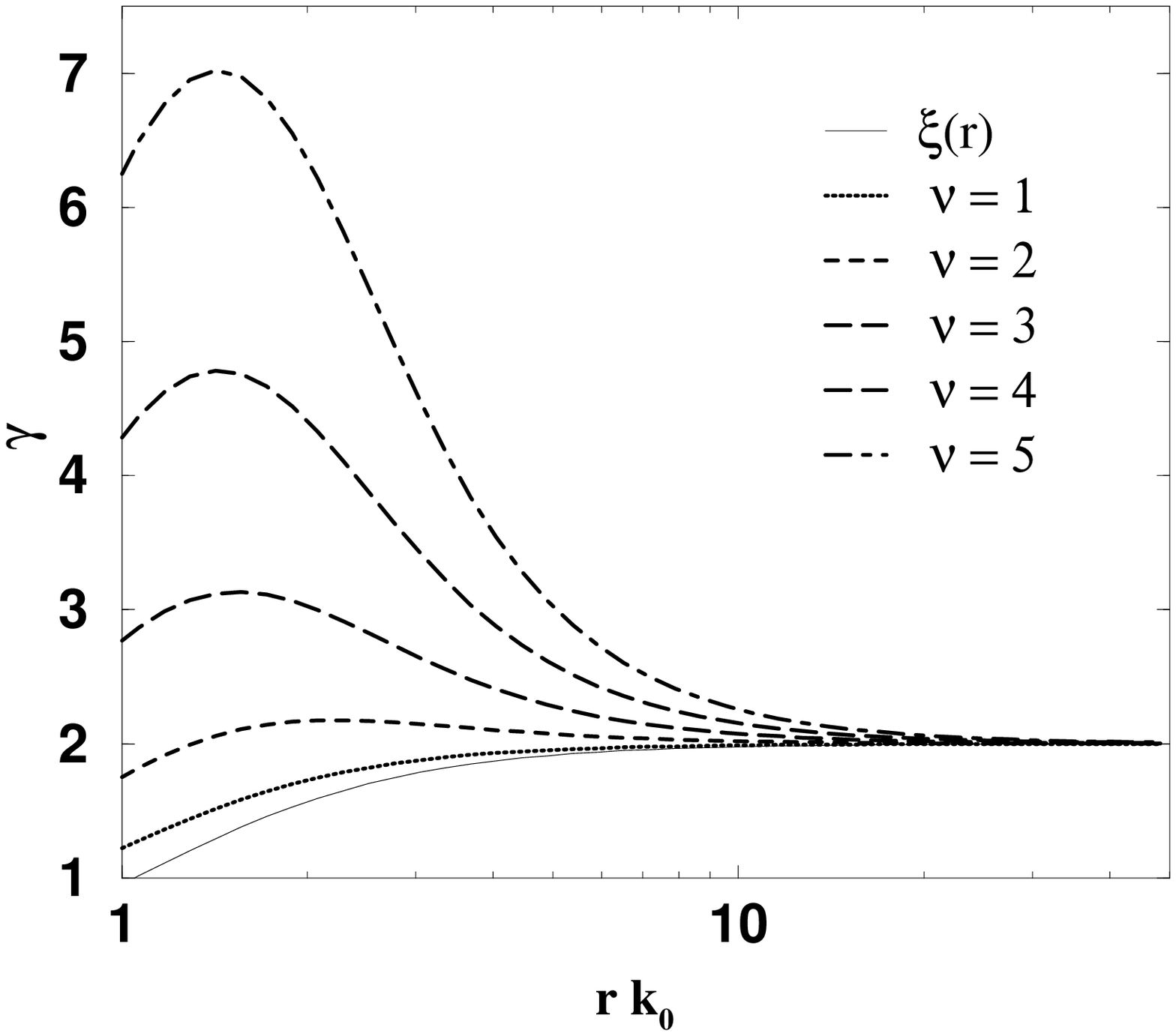]
{\label{fig3}
The behavior of   $\gamma_\nu(r)$ is shown for different values of the
threshold $\nu$ for the correlation function shown in Fig.2.
Clearly $\gamma_\nu$ is strongly scale dependent on all scales where
$\xi_\nu  \gtapprox  1$, this is $r< 1$ in our units. }

\clearpage

\plotone{fig1.eps}

\clearpage

\plotone{fig2.eps}

\clearpage

\plotone{fig3.eps}


\begin{thebibliography}{}


\bibitem{Bahcall} Bahcall N. and Soneira R., Astron. Astrophys. 1983  
	270, 20  

\bibitem{Bardeen} Bardeen J.,  Bond J.R., Kaiser  N. and
	Szalay A.,1986,  ApJ  304, 15 

\bibitem{Cappi} Benoist C.et al.,1996 ApJ  472, 452 
 
\bibitem{Coles} Coles P., 1986, MNRAS 222, 9p  
	 
\bibitem{Feller} Feller W.,1965   An Introduction to Probability Theory and its
	Applications, vol. 2, ed. J. Wiley\& Sons, (New York).


\bibitem{Perez}Gaite J., Dom\'inguez  A.  and P\'erez-Mercader J. 
	1999, ApJ  522, L5 
	
\bibitem{Kaiser} Kaiser N.,  1984, ApJ, 
	  284, L9 

\bibitem{ma-94} Ma  K.S. 1994   Modern Theory of Critical Phenomena, 
	ed. Addison-Wesley  

\bibitem{Pee80} Peebles J.P.E.,1980   The Large Scale Structure of the universe
	Princeton Universuty Press  

\bibitem{Pie87}Pietronero  L., 1987 Physica A  144, 257 


\bibitem{Politzer} Politzer  H.D. \& Wise  M.B. 
	1984, ApJ, 285, L1


\bibitem{Slmp98} Sylos Labini F.,  Montuori M. and
	  Pietronero L., 1998, Phys.Rep,  293, 66  
	 	 
\bibitem{Wax} Rice  S.O. 1954, in   Noise and Stochastic Processes,
	ed. N. Wax, Dover Publications, (New York).


\bibitem{Vanm}   Vanmarcke E., 1983  Random Fields, MIT press (Cambridge
	MA)


\end{thebibliography}
\end{document}